\documentclass[12pt,reqno]{amsart}                
\usepackage{amssymb}
\usepackage{mathrsfs}

\newtheorem{theorem}{Theorem}

\begin{document}

\title[On the game interpretation of a shadow price]{On the game interpretation of a shadow price process in utility maximization problems under transaction costs}

\author{Dmitry B. Rokhlin}
%\address{D.B. Rokhlin, Faculty of Mathematics, Mechanics and Computer Sciences, Southern Federal University, 
%Mil'chakova str., 8a, 344090, Rostov-on-Don, Russia}
%\email{rokhlin@math.rsu.ru}  
\newcommand{\esssup}{{\rm ess\,sup}}
\newcommand{\LIM}{{\rm LIM\,}}
\renewcommand{\theequation}{\thesection.\arabic{equation}}
\thispagestyle{empty}

\begin{abstract}
To any utility maximization problem under transaction costs one can assign a frictionless model with a price process $S^*$, lying in the bid/ask price interval $[\underline S, \overline S]$. Such process $S^*$ is called a \emph{shadow price} if it provides the same optimal utility value as in the original model with bid-ask spread. 

We call $S^*$ a \emph{generalized shadow price} if the above property is true for the \emph{relaxed} utility function in the frictionless model. This relaxation is defined as the lower semicontinuous envelope of the original utility, considered as a function on the set $[\underline S, \overline S]$, equipped with some natural weak topology. 
We prove the existence of a generalized shadow price under rather weak assumptions and mark its relation to a saddle point of the trader/market zero-sum game, determined by the relaxed utility function. 
The relation of the notion of a shadow price to its generalization is illustrated by several examples. Also, we briefly discuss the interpretation of shadow prices via Lagrange duality.
\end{abstract}

\maketitle

\section{Introduction}
\setcounter{equation}{0}

A possible approach to the analysis of optimization problems under transaction costs consists in their reduction to the correspondent problems in frictionless models. The main point of this approach is to determine a frictionless price process $S^*$, called a shadow price, lying in the bid/ask price interval $[\underline S, \overline S]$ and ensuring the same optimal utility value. This method was successfully applied to some continuous time portfolio optimization problems in the recent papers \cite{KalMuh-Kar10}, \cite{GerMuh-KarSch011a}, \cite{GerMuh-KarSch011b}, \cite{GerGuaMuh-KarSch011}. Previously in the same context a shadow price process with such interpretation explicitly appeared in \cite{Low02}.

In discrete time setting for the case of finite probability space the existence of a shadow price in an investment/consumption optimization problem was established in \cite{KalMuh-Kar11}. Inspired by this result, we consider an optimal investment problem in discrete time model over general probability space. It should be mentioned that very recently in the paper \cite{BenCamKalMuh11} the existence of a shadow price process was established in general multi-currency continuous time market models under short selling constraints.

The main feature of the present paper is the game interpretation of a shadow price process. As it was was mentioned in the cited papers, a shadow price can be interpreted as a least favourable frictionless price from trader's point of view. So, it is natural to consider a trader/market zero-sum game determined by trader's utility $\Psi(S,\gamma)$, regarded as a function of frictionless price process $S$ and an investment strategy $\gamma$. Moreover, one can expect that a pair $(\gamma^*,S^*)$, composed of an optimal strategy $\gamma^*$ and a shadow price $S^*$, corresponds to a saddle point of $\Psi$.

However, an application of customary minimax theorems (see \cite{Sim95}) is not straightforward. Firstly, usually $\Psi$ is not convex or quasiconvex in $S$. Secondly, in general it is not lower semicontinuous in a topology, ensuring the compactness of the set $[\underline S,\overline S]$. To overcome at least the second difficulty, for each $\gamma$ we pass to the lower semicontinuous envelope $\widehat\Psi$ of $\Psi$ in some natural weak topology on $[\underline S,\overline S]$, and introduce the corresponding notion of a generalized shadow price process $S^*$.

The method, involving a consideration of the lower semicontinuous envelope (relaxation) of the objective functional is  extensively used in analysis of variational problems \cite{EkeTem99}, \cite{AttButMic06}, \cite{DalM93}. 
In the present context it appears that the relaxed problem fits nicely into the framework of the intersection theorem, proved by Ha \cite{Ha80} (see Theorem \ref{th:2} below).
Applying this result, in Section \ref{sec:2} we establish the existence of a generalized shadow price and the minimax property of $\widehat\Psi$ under rather weak assumptions  (Theorem \ref{th:1}). Moreover, if there exists an optimal solution $\gamma^*$ of the original utility maximization problem, then a pair $(\gamma^*,S^*)$, where  $S^*$ is a generalized shadow price, is exactly the strategic saddle point of the game, determined by the relaxed utility function $\widehat\Psi(S,\gamma)$ (Theorem \ref{th:3}). 

Thus, the advantage of passing to the relaxed problem is twofold: (1) the existence of a generalized shadow price process $S^*$ is guaranteed under weak assumptions, (2) the relaxed utility $\widehat\Psi$ has nice minimax and saddle-point properties. 
 
The relation of the notion of a shadow price to its generalization is illustrated  by several examples in Section \ref{sec:3}.
If the original utility function $\Psi(S,\gamma)$ is already lower semicontinuous in $S$ in an appropriate topology, the proposed approach gives the existence of a shadow price (Examples 1 and 2). Another interesting case appears when $\widehat\Psi\neq\Psi$ but it is still possible to give a convenient analytical description of the saddle points $(S^*,\gamma^*)$ of $\widehat\Psi$. If $\gamma^*$ is
an optimal solution of the original utility maximization problem under transaction costs, $\Psi(S^*,\gamma^*)=\widehat\Psi(S^*,\gamma^*)$ and the optimality of $\gamma^*$ for the functions
$\gamma\mapsto\Psi(S^*,\gamma)$, $\gamma\mapsto\widehat\Psi(S^*,\gamma)$ is characterized by identical conditions,
then a generalized shadow price $S^*$ is in fact a shadow price (Example 3). 

Furthemore, we give an example of two-step model on a countable probability space with linear utility functional such that a generalized shadow price exists and a shadow price is not (Example 4). Independently an example of the same nature in three-step model was constructed in \cite{BenCamKalMuh11}. In spite of the nonlinearity of the objective functional, the advantage of the latter example is the use of logarithmic utility, while in Example 4 the utility contains a Banach limit. We find it interesting to test our approach on the example of \cite{BenCamKalMuh11}. It appears that the ''unsuccessful candidate'' for a shadow price, mentioned in \cite{BenCamKalMuh11}, is a generalized shadow price (see Example 5 of the present paper). 

It is worth mentioning that usually a ''shadow'' or ''equilibrium'' resource prices are associated with an optimal solution of the Lagrange (or Fenchel) dual problem. In Section \ref{sec:4} we trace this connection in the problem under consideration, confining ourselves to the case of finite probability space. We show that a shadow price is equal to the relation of equilibrium prices of stock and bond. The related calculations indicate quite explicitly that the zero duality gap and the solvability of the Lagrange dual problem immediately imply the existence of a shadow price process. This point seems promising for generalizations, concerning the existence of a shadow price. However, Examples 4 and 5 show that this way is not so easy in the infinite-dimensional setting. See also the comments, concerning the papers \cite{CviKar96}, \cite{CviWan01}, in the introductory section of \cite{BenCamKalMuh11}. 

\section{Main result} \label{sec:2}
\setcounter{equation}{0}
Consider a trader, who can distribute his wealth between a bond with zero interest rate (and price 1) and a risky asset (stock). As usual, he acts in random setting, described by a probability space $(\Omega,\mathcal F,\mathsf P)$ endowed with discrete-time filtration $\mathbb F=(\mathcal F_t)_{t=-1}^T$, $\mathcal F_{-1}=\{\emptyset,\Omega\}$. The stock can be sold at the bid price $\underline S_t$ and purchased at the ask price $\overline S_t$ at a time moment $t$. It is assumed that $0<\underline S_t\le\overline S_t$ and the processes $\underline S$, $\overline S$ are $\mathbb F$-adapted. A trading strategy is determined by an $\mathbb F$-adapted portfolio process $(\beta_t,\gamma_t)_{t=-1}^T$, consisting of $\beta_t$ untis of bond (or cash) and $\gamma_t$ units of stock. A trading strategy is called \emph{self-financing} if any change in risky position is compensated by the cash flow:
$$ \Delta\beta_t=\underline S_t(\Delta\gamma_t)^- - \overline S_t(\Delta\gamma_t)^+,\ \ t=0,\dots,T,$$
where $\Delta a_t=a_t-a_{t-1}$, $x^+=\max\{x,0\}$, $x^-=\max\{-x,0\}$. To fix the values $\beta_{-1}$, $\gamma_{-1}$ we assume that the trader starts from one unit of bond: $\beta_{-1}=1$, $\gamma_{-1}=0$. Moreover, at the terminal date $T$ the asset holdings are converted to cash: $\gamma_T=0$. Hence, trader's terminal wealth is given by
\begin{equation} \label{eq:2.1}
X_T(\gamma)=1+\sum_{t=0}^T (\underline S_t (\Delta\gamma_t)^- - \overline S_t (\Delta\gamma_t)^+),\ \ \ 
\gamma_{-1}=0,\ \gamma_T=0.
\end{equation}
For the frictionless model ($S=\underline S=\overline S$) this formula shapes to the customary form:
$$ X_T(\gamma)=1+(\gamma\circ S)_T:=1+\sum_{t=1}^T\gamma_{t-1}\Delta S_t.$$

Denote by $L^0(\mathcal F_t)$ the set of equivalence classes of $\mathsf P$-a.s. equal $\mathcal F_t$-measurable real-valued random variables. The sets $L^p(\mathcal F_t)$, $1\le p<\infty$ and $L^\infty(\mathcal F_t)$ consist of $p$-th power $\mathsf P$-integrable and $\mathsf P$-essentially bounded elements of $L^0(\mathcal F_t)$ respectively. We equip $L^0$ with the topology of convergence in probability, induced by the metric
\begin{equation} \label{eq:2.2}
\rho(f,g)=\mathsf E\frac{|f-g|}{1+|f-g|}. 
\end{equation}

Unless otherwise stated, the sets $L^p$, $p\in [1,\infty)$; $L^\infty$ are considered as Banach spaces with the norms
$$ \|f\|_p=\left(\mathsf E|f|^p\right)^{1/p},\ \ \ \ \|f\|_\infty=\esssup\, |f|.$$

We consider two possible choices of spaces, containing the portfolio strategies $\gamma$: $\gamma_t\in L^s(\mathcal F_t)$, $s\in\{0,\infty\}$. However, in each case we equip $L^s(\mathcal F_t)$ with the  topology $\tau_t$ of convergence in probability. Denote by $\mathfrak F$ the vector space $\prod_{t=0}^{T-1} L^s(\mathcal F_t)$ with the product topology $\tau=\prod_{t=0}^{T-1} \tau_t$  and let $\mathcal Y$ be a convex subset of $\mathfrak F$. Since the values $\gamma_{-1}=0$, $\gamma_T=0$ are fixed, in what follows $\gamma$ is considered as an element of $\mathfrak F$ (except for Section \ref{sec:4}). We allow portfolio constraints of the form $\gamma\in\mathcal Y$.
% and let $\mathfrak F$ be a closed convex subset of $\mathfrak F$.  We allow the portfolio constraints of the form $\gamma\in\mathfrak F$.

Assume that $\underline S_t,\overline S_t\in L^q(\mathcal F_t)$ for some $q\in [1,\infty]$. Put $\tau^w_t=\sigma(L^q(\mathcal F_t),L^p(\mathcal F_t))$, where $1/p+1/q=1$. So, $\tau_t^w$ is the weak topology of $L^q$ for $q\in [1,\infty)$ and the weak-star topology of $L^\infty$. In any case the set 
$$ [\underline S_t,\overline S_t]=\{S_t\in L^q(\mathcal F_t): \underline S_t\le S_t\le \overline S_t\}$$
is $\tau_t^w$-compact. Since the closedness of $[\underline S_t,\overline S_t]$ is clear, this assertion follows from the $\tau_t^w$-compactness of the unit ball for $q\in (1,\infty]$ and the uniform integrability of $[\underline S_t,\overline S_t]$ for $q=1$. Denote by $\mathfrak E$ the vector space $\prod_{t=0}^T L^q(\mathcal F_t)$ with the product topology $\tau^w=\prod_{t=0}^T\tau_t^w$ and put 
$$\mathcal X=[\underline S,\overline S]:=\prod_{t=0}^T [\underline S_t,\overline S_t].$$

A functional
$$ \Phi: L^r(\mathcal F_T)\mapsto [-\infty,\infty],\ \ r=\min\{s,q\}.$$
$\Phi$ is called \emph{monotone} if $\Phi(X)\ge \Phi(Y)$ whenever $X\ge Y$, $X,Y\in L^r(\mathcal F_T)$ and \emph{quasiconcave} if 
$$ \Phi(\alpha_1 X+\alpha_2 Y)\ge\min\{\Phi(X),\Phi(Y)\} $$
for all $X, Y\in L^r(\mathcal F_T)$, $\alpha_1+\alpha_2=1$, $\alpha_i\ge 0$. It is easy to see that $\Phi$ is quasiconcave iff the upper level sets $\{X\in L^r(\mathcal F_T): \Phi(X)>\beta\}$ are convex for all $\beta\in\mathbb R$. We admit that trader's preferences are represented by a monotone quasiconcave functional $\Phi$. Recently such framework attracted a considerable attention in connection with quasiconvex risk measures \cite{CerMacMarMon11}, \cite{FriGia11}.

The optimal value of the utility maximization problem under transaction costs and portfolio constraints, represented by $\mathcal Y$, is defined as follows
\begin{equation} \label{eq:2.3}
 \lambda=\sup\{\Phi(X_T(\gamma)):\gamma\in\mathcal Y\}.
\end{equation}
Note that $X_T(\gamma)\in L^r(\mathcal F_T)$ under the above notation.

Along with (\ref{eq:2.3}) consider the optimization problem in a frictionless model, where the stock price is given by an adapted process $S\in [\underline S,\overline S]$: 
\begin{equation} \label{eq:2.4}
 \mu_S=\sup\{\Phi(1+(\gamma\circ S)_T):\gamma\in\mathcal Y\}.
\end{equation}
Following \cite{KalMuh-Kar11} we call an adapted process $S\in [\underline S,\overline S]$ a \emph{shadow price} if $\mu_S=\lambda$.

We are going to introduce a modification of the last notion. Put 
$$ \Psi(S,\gamma)=\Phi(1+(\gamma\circ S)_T)$$
and denote by $\widehat\Psi(\cdot,\gamma)$ the $\tau^w$-lower semicontinuous envelope (relaxation) of $\Psi(\cdot,\gamma)$ as a function on $[\underline S,\overline S]$ (see \cite{PapKyr09}, Definition 2.1.13):
$$ \widehat\Psi(S,\gamma)=\sup_{V\in\mathcal N(S)} \inf_{S'\in V}\Psi(S',\gamma),$$
where $\mathcal N(S)$ is a local base of the topology $\tau^w$, restricted to $[\underline S,\overline S]$. 
As is known (see \cite{PapKyr09}, Proposition 2.1.15), $\widehat\Psi(\cdot,\gamma)$ is the largest $\tau^w$-lower semicontinuous function majorized by $\Psi(\cdot,\gamma)$. 
Note that 
\begin{equation} \label{eq:2.5}
\Phi(X_T(\gamma))\le\widehat \Psi(S,\gamma)\le\Psi(S,\gamma),\ \ \ S\in [\underline S,\overline S]
\end{equation}
since $X_T(\gamma)\le 1+(\gamma\circ S')_T$, $\underline S\le S'\le\overline S$ and $\Phi$ is monotone.

Consider instead of (\ref{eq:2.4}) the optimization problem for the relaxed functional $\widehat\Psi$:
\begin{equation} \label{eq:2.6}
\widehat\mu_S=\sup\{\widehat\Psi(S,\gamma):\gamma\in \mathcal Y\}.
\end{equation}
We call $S$ a \emph{generalized shadow price} if $\widehat\mu_S=\lambda$. 

Looking at (\ref{eq:2.5}), we immediately conclude that any shadow price $S^*$ is a generalized shadow price: 
$$\lambda=\sup_{\gamma\in \mathcal Y}\Phi(X_T(\gamma))\le\sup_{\gamma\in \mathcal Y}\widehat \Psi(S^*,\gamma)=\widehat\mu_{S^*}\le\sup_{\gamma\in \mathcal Y}\Psi(S^*,\gamma)=\lambda.$$

If $\Phi$ is quaisiconcave then $\Psi(S,\cdot)$, $\widehat\Psi(S,\cdot)$ are quasiconcave as well. Indeed, for $\alpha_1+\alpha_2=1$, $\alpha_i\ge 0$ and $\gamma^i\in\mathcal Y$ we have
\begin{eqnarray*}
\Psi(S,\alpha_1\gamma^1+\alpha_2\gamma^2)=\Phi(\alpha_1 (1+(\gamma^1\circ S)_T)+\alpha_2 (1+(\gamma^1\circ S)_T))\\ \ge\min_{i=1,2}\Phi(1+(\gamma^i\circ S)_T)=\min_{i=1,2}\Psi(S,\gamma^i).
\end{eqnarray*}
Let $\widehat\Psi(S,\gamma^i)>\beta$. %Since $\widehat\Psi(\cdot,\gamma)$ is lower semicontinuous, the sets $\{S'\in\mathcal X:\widehat\Psi(S',\gamma^i)>\alpha\}$ are $\tau^w$-open. 
Take $V^i\in\mathcal N(S)$ such that $\inf_{S'\in V^i}\Psi(S',\gamma^i)>\beta$ and put $V=V^1\cap V^2$. The inequality
$$\widehat\Psi(S,\alpha_1\gamma^1+\alpha_2\gamma^2)\ge\inf_{S'\in V}\Psi(S',\alpha_1\gamma^1+\alpha_2\gamma^2) \ge\min_{i=1,2}\inf_{S'\in V}\Psi(S',\gamma^i)>\beta$$
means that the upper level sets $\{\gamma\in\mathcal Y:\widehat\Psi(S,\gamma)>\beta\}$ are convex.

Now we state the main result of the present paper.
 
\begin{theorem} \label{th:1}
Let $\Phi$ be monotone and quasiconcave, $\underline S_t,\overline S_t\in L^q(\mathcal F_t)$, $t=0,\dots,T$ for some $q\in [1,\infty]$. Then there exists a generalized shadow price $S^*\in[\underline S,\overline S]$ and the following minimax relations hold true:
\begin{eqnarray} \label{eq:2.7}
\lambda=\sup_{\gamma\in\mathcal Y}\Phi(X_T(\gamma))=\sup_{\gamma\in\mathcal Y}\inf_{S\in [\underline S,\overline S]}\widehat\Psi(S,\gamma)=\sup_{\gamma\in\mathcal Y}\inf_{S\in [\underline S,\overline S]}\Psi(S,\gamma)\\
=\inf_{S\in [\underline S,\overline S]}\sup_{\gamma\in\mathcal Y}\widehat\Psi(S,\gamma)\nonumber 
=\sup_{\gamma\in\mathcal Y}\widehat\Psi(S^*,\gamma)=\widehat\mu_{S^*}.
\end{eqnarray}
\end{theorem} 

In fact, Theorem \ref{th:1} is a direct consequence of the following intersection theorem (\cite{Ha80}, Theorem 3). We formulate it in a slightly weaker form.
\begin{theorem}[Ha, 1980] \label{th:2}
Let $\mathfrak E$, $\mathfrak F$ be Hausdorff topological vector spaces, $\mathcal X\subset\mathfrak E$ be a convex compact set, $\mathcal Y\subset\mathfrak F$ be a convex set. Let $B\subset A\subset\mathcal X\times\mathcal Y$ be subsets such that
\begin{itemize}
\item[(a)] for each $y\in\mathcal Y$ the set $\{x\in\mathcal X:(x,y)\in A\}$ is closed;
\item[(b)] for each $x\in\mathcal X$ the set $\{y\in\mathcal Y:(x,y)\not\in A\}$ is convex;
\item[(c)] $B$ is closed in $\mathcal X\times\mathcal Y$ and for each $y\in\mathcal Y$ the set $\{x\in\mathcal X:(x,y)\in B\}$ is nonempty and convex.
\end{itemize}
Then there exists a point $x^*\in\mathcal X$ such that $\{x^*\}\times\mathcal Y\subset A$.
\end{theorem} 

\emph{Proof of Theorem \ref{th:1}}. The topological vector spaces $(\mathfrak E,\tau^w)$, $(\mathfrak F,\tau)$ and sets $\mathcal X=[\underline S,\overline S]$, $\mathcal Y$, introduced above, satisfy the conditions of Theorem \ref{th:2}. Put
$$ A=\{(S,\gamma)\in \mathcal X\times\mathcal Y:\widehat\Psi(S,\gamma)\le\lambda\}.$$
Condition (a) of Theorem \ref{th:2} is satisfied since $\widehat\Psi(\cdot,\gamma)$ is $\tau^w$-lower semicontinuous and the validity of (b) follows from the quasiconcavity of $\widehat\Psi(S,\cdot)$:
$$ \{\gamma\in\mathcal Y:(S,\gamma)\not\in A\}=\{\gamma\in\mathcal Y:\widehat\Psi(S,\gamma)>\lambda\}.$$

Furthermore, consider the set-valued mapping $\widehat B$ from $\mathcal Y$ to the power set of $\mathcal X$, defined as follows
%\begin{equation} \label{eq:2.7}
$$\widehat B(\gamma)=\left(\{\underline S_t\}I_{\{\Delta\gamma_t<0\}}+\{\overline S_t\}I_{\{\Delta\gamma_t>0\}}
+[\underline S_t,\overline S_t] I_{\{\Delta\gamma_t=0\}}\right)_{t=0}^T$$
%\end{equation}
and denote by $B$ the graph of $\widehat B$: 
$$B=\{(S,\gamma)\in\mathcal X\times\mathcal Y: S\in\widehat B(\gamma)\}.$$ 
For $(S,\gamma)\in B$ we have %$1+(\gamma\circ S)_T=X^\gamma_T$
\begin{equation} \label{eq:2.8}
X_T(\gamma)=1+\sum_{t=0}^T (\underline S_t (\Delta\gamma_t)^- - \overline S_t (\Delta\gamma_t)^+)
=1-\sum_{t=0}^T S_t \Delta\gamma_t=1+(\gamma\circ S)_T
\end{equation}
and $\Phi(X_T(\gamma))=\Psi(S,\gamma)\le\lambda$. Thus, $\widehat\Psi(S,\gamma)\le\lambda$ and $B\subset A$.

We claim that $B$ satisfies condition (c) of Theorem \ref{th:2}. Cleary, the sets $\{S\in\mathcal X:(S,\gamma)\in B\}=\widehat B(\gamma)$ are nonempty and convex. It remains to prove that $B$ is closed in $\mathcal X\times\mathcal Y$. Let $(S,\gamma)\in\mathcal X\times\mathcal Y$ lie in the closure of $B$ in the product topology $\tau^w\times\tau$, restricted to $\mathcal X\times\mathcal Y$. To prove that $(S,\gamma)\in B$ it is sufficient to show that
\begin{equation} \label{eq:2.9}
S_t I_{\{\Delta\gamma_t\neq 0\}}=\underline S_t I_{\{\Delta\gamma_t<0\}}+\overline S_t I_{\{\Delta\gamma_t>0\}},\ \ t=0,\dots, T.
\end{equation}

For any $t\in\{0,\dots,T\}$, $n\in\mathbb N$ and $g_t\in L^\infty(\mathcal F_t)$ there exist $\gamma^n\in\mathcal Y$ and $S^n\in [\underline S,\overline S]$ of the form
$$S^n=\underline S I_{\{\Delta\gamma^n<0\}}+\overline S I_{\{\Delta\gamma^n>0\}}+\widehat S^n I_{\{\Delta\gamma^n=0\}},\ \ \underline S\le \widehat S^n\le \overline S$$
such that $\rho(\gamma_t^n,\gamma_t)<1/n$, $|\mathsf E(S_t^n-S_t)g_t I_{\{\Delta\gamma_t\neq 0\}}|<1/n$, where $\rho$ is defined by (\ref{eq:2.2}). Passing to subsequences (still denoted by $\gamma^n_t$, $S^n_t$), we may assume that $\gamma_t^n\to \gamma_t$ $\mathsf P$-a.s. Here $\gamma^n_t$, $\gamma_t$ are understood as functions, taken from the correspondent equivalence class. 

On the set $\{\Delta\gamma_t\neq 0\}$ we have 
$$I_{\{\Delta\gamma_t^n<0\}}\to I_{\{\Delta\gamma_t<0\}}, \ \
  I_{\{\Delta\gamma_t^n>0\}}\to I_{\{\Delta\gamma_t>0\}}, \ \
  I_{\{\Delta\gamma_t^n=0\}}\to 0\ \ \mathsf P\textnormal{-a.s.}$$
From the the dominated convergence theorem it follows that 
$$  \lim_{n\to\infty}\mathsf E\left(g_t (\underline S_t I_{\{\Delta\gamma_t^n<0\}}+\overline S_t I_{\{\Delta\gamma_t^n>0\}})I_{\{\Delta\gamma_t\neq 0\}}\right)= \mathsf E(g_t (\underline S_t I_{\{\Delta\gamma_t<0\}}+\overline S_t I_{\{\Delta\gamma_t>0\}}),$$
$$ \left|\mathsf E\left(g_t \widehat S_t^n I_{\{\Delta\gamma_t^n=0\}}I_{\{\Delta\gamma_t\neq 0\}}\right)\right|\le
\mathsf E\left(|g_t| \overline S_t I_{\{\Delta\gamma_t^n=0\}}I_{\{\Delta\gamma_t\neq 0\}}\right)\to 0. $$ 
Hence, 
$$ \mathsf E\left(g_t S_t I_{\{\Delta\gamma_t\neq 0\}}\right)=\lim_{n\to\infty}\mathsf E\left(g_t S_t^n I_{\{\Delta\gamma_t\neq 0\}}\right)=
 \mathsf E\left(g_t (\underline S_t I_{\{\Delta\gamma_t<0\}}+\overline S_t I_{\{\Delta\gamma_t>0\}})\right)$$
for any $g_t\in L^\infty(\mathcal F_t)$ and (\ref{eq:2.9}) is satisfied. 

Now we can apply Theorem \ref{th:2} and take an element $S^*\in\mathcal X$ such that $\{S^*\}\times\mathcal Y\subset A$. That is, $\widehat\Psi(S^*,\gamma)\le \lambda$ for all $\gamma\in\mathcal Y$ and
$$\widehat\mu_{S^*}=\sup\{\widehat\Psi(S^*,\gamma):\gamma\in\mathcal Y\}\le\lambda.$$
The reverse inequality is clear. Thus, $S^*$ is a generalized shadow price.

The equalities in the first line in (\ref{eq:2.7}) follow from (\ref{eq:2.5}) and (\ref{eq:2.8}). Furthermore, the inequalities 
$$\sup_{\gamma\in\mathcal Y}\inf_{S\in [\underline S,\overline S]}\widehat\Psi(S,\gamma)\\
\le\inf_{S\in [\underline S,\overline S]}\sup_{\gamma\in\mathcal Y}\widehat\Psi(S,\gamma)\nonumber 
\le\sup_{\gamma\in\mathcal Y}\widehat\Psi(S^*,\gamma)$$
are evident and the equality $\lambda=\sup_{\gamma\in\mathcal Y}\widehat\Psi(S^*,\gamma)$ is already proved. \qed

In the context of duality theory (see e.g. \cite{Roc74}, section 1) the problems
\begin{eqnarray} \label{eq:2.10}
\textnormal{maximize}\ \   f(\gamma) &=& \inf_{S\in [\underline S,\overline S]}\widehat\Psi(S,\gamma) \ \ \textnormal{over\ all}\ \gamma\in\mathcal Y,\nonumber\\
\textnormal{minimize} \ \  g(S) &=& \sup_{\gamma\in\mathcal Y}\widehat\Psi(S,\gamma) \ \ \textnormal{over\ all}\ S\in [\underline S,\overline S]
\end{eqnarray}
are said to be \emph{dual} to each other and the common value (\ref{eq:2.7}) is called the \emph{saddle-value} of $\widehat\Psi$. The first of these problems coincides with (\ref{eq:2.3}).

The function $g$ is lower semicontinuous as the pointwise supremum of a family of lower semicontinuous
functions and the set $[\underline S,\overline S]$ is compact. Hence, the dual problem (\ref{eq:2.10}) is solvable. The equality
$$ g(S^*)=\sup_{\gamma\in\mathcal Y}\widehat\Psi(S^*,\gamma)=\inf_{S\in [\underline S,\overline S]}\sup_{\gamma\in\mathcal Y}\widehat\Psi(S,\gamma)=\lambda$$
%for a generalized shadow price $S^*$ 
shows that generalized shadow prices are exactly the solutions of (\ref{eq:2.10}).

\begin{theorem} \label{th:3}
A pair $(S^*,\gamma^*)\in [\underline S,\overline S]\times\mathcal Y$ is a saddle point of the relaxed utility function $\widehat\Psi$:
\begin{equation} \label{eq:2.11}
 \widehat\Psi(S^*,\gamma)\le\widehat\Psi(S^*,\gamma^*)\le\widehat\Psi(S,\gamma^*),\ \ \ 
(S,\gamma)\in [\underline S,\overline S]\times\mathcal Y,
\end{equation}
if and only if $\gamma^*$ is an optimal solution of (\ref{eq:2.3}): $\Phi(X_T(\gamma^*))=\lambda$ and $S^*$ is a generalized shadow price.
\end{theorem}

\emph{Proof}. 
%A pair $(S^*,\gamma^*)\in\mathcal X\times\mathfrak F$ is called a \emph{saddle point} of $\widehat\Psi$ if
%$$  \widehat\Psi(S^*,\gamma)\le\widehat\Psi(S^*,\gamma^*)\le\widehat\Psi(S,\gamma^*),\ \ \ 
%(S,\gamma)\in\mathcal X\times\mathfrak F. $$
Condition (\ref{eq:2.11}) can be reformulated as follows:
\begin{equation} \label{eq:2.12}
g(S^*)=\widehat\Psi(S^*,\gamma^*)=f(\gamma^*).
\end{equation}
Since $g(S)\ge f(\gamma)$, $(S,\gamma)\in [\underline S,\overline S]\times\mathcal Y$ it follows that if $(S^*,\gamma^*)$ is a saddle point of $\widehat\Psi$ then $\gamma^*$ is an optimal solution of (\ref{eq:2.3}) and $S^*$ is an optimal solution of (\ref{eq:2.10}) (or, equivalently, a generalized shadow price). Conversely, if $\gamma^*$ is an optimal solution of (\ref{eq:2.3}) and $S^*$ is a generalized shadow price then
$$ \lambda=g(S^*)\ge\widehat\Psi(S^*,\gamma^*)\ge f(\gamma^*)=\lambda.$$
Thus, (\ref{eq:2.12}) holds true and $(S^*,\gamma^*)$ is a saddle point of $\widehat\Psi$. \qed

The above arguments show that the existence of an optimal solution of (\ref{eq:2.3}) is equivalent to the existence of a saddle point of the relaxed utility function $\widehat\Psi$.

\section{Examples} \label{sec:3}
\setcounter{equation}{0}
In the first two examples given below $\Psi(\cdot,\gamma)$ is $\tau^w$-lower semicontinuous on $[\underline S,\overline S]$ and, hence, there exists a shadow price. Note that in these examples the topological space $([\underline S,\overline S],\tau^w)$ is first countable and it is enough to show that
\begin{equation} \label{eq:3.1}
 \Psi(S,\gamma)\le\liminf_{n\to\infty}\Psi(S^n,\gamma),\ \ S\in [\underline S,\overline S]
\end{equation}
for any sequence $S^n\in [\underline S,\overline S]$, converging to $S$ in $\tau^w$, to check $\tau^w$-lower semicontinuity of $\Psi(\cdot,\gamma)$.

In Example 3 the relaxation $\widehat\Psi$ does not coincide with $\Psi$ but $S^*$ is a shadow price iff it is a generalized shadow price. Examples 4 and 5 (the last one is borrowed from \cite{BenCamKalMuh11}) show that even in the case of countable probability space a shadow price need not exist, while the existence of a generalized shadow price is ensured by Theorem \ref{th:1}. 

In all examples the utility functional $\Phi$ is concave and $\mathcal Y=\mathfrak F$.

\emph{Example 1.} Let $\Omega$ be finite and let $\mathsf P$ be striclty positive on the atoms of $\mathcal F_T$. Consider a monotone concave function $U:\mathbb R\mapsto [-\infty,\infty)$  such that $U$ is finite (and hence continuous) on the open half-line $(0,\infty)$ and $U(x)=-\infty,\ x\in (-\infty,0]$. We look for a shadow price in the optimization problem
$$\textnormal{maximize}\ \ \mathsf E U(X_T(\gamma))\ \ \textnormal{over all}\ \ \gamma\in\prod_{t=0}^{T-1} L^0(\mathcal F_t)$$
where $X_T(\gamma)$ is defined by (\ref{eq:2.1}). The choice of $q$ does not affect anything: $q=1$ for instance.

If $\mathcal F_t$ is generated by the partition $(D_t^i)_{i=1}^{m_t}$ then $L^1(\mathcal F_t)$ is an $m_t$-dimensional space. Put $f_t^i=f_t(\omega)$, $\omega\in D_t^i$ for $f_t\in L^1(\mathcal F_t)$. All Hausdorff vector topologies on a finite dimensional space coincide (\cite{AliBod06}, Theorem 5.21). Thus, we can assume that $\tau_t^w$ is the topology of pointwise convergence with a local base at zero generated by the sets
$$ \{f\in L^1(\mathcal F_t): |f_t^i|<1/n\},\ \ i=1,\dots,m_t,\ \ n\in\mathbb N.$$

To show that $\Psi(\cdot,\gamma)$ is lower semicontinuous in the product topology $\tau^w=\prod_{t=0}^T\tau^w_t$ it is enough to check that (\ref{eq:3.1}) is true when $S^n_t\to S_t$ pointwise.

If $1+(\gamma\circ S)_T(\omega)>0$ for all $t$, $\omega$ then the same is true for $1+(\gamma\circ S^n)_T(\omega)$ for sufficiently large $n$. It follows that
$$ \lim_{n\to\infty}\Psi(S^n,\gamma)=\lim_{n\to\infty}\mathsf E U(1+(\gamma\circ S^n)_T)=\mathsf E U(1+(\gamma\circ S)_T)=\Psi(S,\gamma).$$
On the other side, if  $1+(\gamma\circ S)_T(\omega)\le 0$ for some $t$, $\omega$ then
$\Psi(S,\gamma)=\Phi(1+(\gamma\circ S)_T)=-\infty$.

The lower semicontinuity of $\Psi(\cdot,\gamma)$ implies the existence of a shadow price. A related result was established in \cite{KalMuh-Kar11}. 

\emph{Example 2.} Let $\Omega$ be countable and let $\mathsf P$ be strictly positive on the atoms of $\mathcal F_T$. Assume that the processes $\underline S$, $\overline S$ are bounded: $\underline S_t, \overline S_t\in L^\infty(\mathcal F_t)$, and consider the optimization problem 
$$\textnormal{maximize}\ \ \mathsf E U(X_T(\gamma))\ \ \textnormal{over all}\ \ \gamma\in\prod_{t=0}^{T-1} L^\infty(\mathcal F_t)$$
with a monotone concave (and hence continuous) function $U:\mathbb R\mapsto\mathbb R$. We put $q=1$.
 
If $\mathcal F_t$ is generated by a partition $(D_t^i)_{i\in J_t}$, $J_t\subset\mathbb N$, then for $f_t\in L^1(\mathcal F_t)$ we put $f_t^i=f_t(\omega)$, $\omega\in D_t^i$. Consider on $L^1(\mathcal F_t)$  the topology $\tau^p_t$ of pointwise convergence with a local base at zero generated by the sets
$$ \{f\in L^1(\mathcal F_t): |f_t^i|<1/n\},\ \ i\in J_t,\ \ n\in\mathbb N.$$
The topologies $\tau^w_t=\sigma(L^1(\mathcal F_t),L^\infty(\mathcal F_t))$, $\tau^p_t$ are different on $L^1(\mathcal F_t)$ if the set $J_t$ is infinite, since $\tau^p$ is first countable and $\tau^w$ is not (see \cite{AliBod06}, Theorem 6.26). Clearly, $\tau^p_t\subset\tau^w_t$. It follows that they coincide on the set $[\underline S_t,\overline S_t]$ which is $\tau^w_t$-compact and $\tau^p_t$-Hausdorff (see \cite{Rud91}, section 3.8). 

Take a sequence $S^n\in [\underline S,\overline S]$, converging to $S$ in the product topology
$\tau^p=\prod_{t=0}^T\tau^p_t$. This amounts to the pointwise convergence $S^n_t\to S_t$. The correspondent sequence $U(1+(\gamma\circ S^n)_T)$ is uniformly bounded and  $\Psi(\cdot,\gamma)$ is $\tau^p$-continuous on $[\underline S,\overline S]$:
$$ \lim_{n\to\infty}\Psi(S^n,\gamma)=\lim_{n\to\infty}\mathsf E U(1+(\gamma\circ S^n)_T)=\mathsf E U(1+(\gamma\circ S)_T)=\Psi(S,\gamma)$$
due to the dominated convergence theorem. This implies the existence of a shadow price.

A counterexample, given in \cite{BenCamKalMuh11} (see Example 5 below), indicates that the assumptions on boundedness of $\underline S_t, \overline S_t$ and finiteness of $U$ cannot be dropped simultaneously.

\emph{Example 3.} Let $T=1$, $\Omega=[0,1]$, $\mathcal F_0=\{\emptyset,\Omega\}$, and let $\mathcal F_1$ be the Borel $\sigma$-algebra of $[0,1]$ with the Lebesgue measure $\mathsf P(d\omega)=d\omega$. Assume that $\underline S_0=\overline S_0=S_0$, $\underline S_1, \overline S_1 \in L^\infty(\mathcal F_1)$ and
$$ \overline S_1-\underline S_1\ge \alpha>0$$
for some real number $\alpha>0$. Consider the optimization problem
%$$\mathsf E U\left(1+\sum_{t=0}^1 \underline S_t (\Delta\gamma_t)^{-}-\overline S_t (\Delta\gamma_t)^{+}\right)=\mathsf 
%\begin{equation} 
\begin{equation} \label{eq:3.2}
\textnormal{maximize}\ \ \mathsf E U(X_1(\gamma))\ \ \textnormal{over all}\ \ \gamma_0\in\mathbb R
\end{equation}
%\end{equation}
%$$\lambda=\sup\{\mathsf EU(X_1^\gamma):\gamma_0\in\mathbb R\},\ \ X_1^\gamma=1+\gamma_0^+ (\underline S_1-S_0)-\gamma_0^- (\overline S_1-S_0)\right)$$
with a monotone concave function $U:\mathbb R\mapsto\mathbb R$. From (\ref{eq:2.1}) we get
$$ X_1(\gamma)=1+\gamma_0^+ (\underline S_1-S_0)-\gamma_0^- (\overline S_1-S_0)$$

Put $s=\infty$, $q=1$. Thus, $[\underline S_1,\overline S_1]$ is considered as a set in $L^1(\mathcal F_1)$ with the weak topology $\tau_1^w$ of $L^1(\mathcal F_1)$. We look for the lower semicontinuous envelope of the functional 
$$ S\mapsto\Psi(S,\gamma)=\mathsf E U(1+\gamma_0(S_1-S_0))$$
defined on the set $\{S_0\}\times [\underline S_1,\overline S_1]$. This problem reduces to relaxation of the integral functional
$$ S_1\mapsto F(S_1)=\int_0^1 \left[U\left(1+\gamma_0 (S_1-S_0)\right)+\delta(S_1|[\underline S_1,\overline S_1])\right]\,d\omega,\ \ S_1\in L^1(\mathcal F_1), $$
where $\delta(x|A)=0$, $x\in A$; $\delta(x|A)=+\infty$, $x\not\in A$ since
$$ F(S_1)=\left\{\begin{array}{cc}
                \Psi(S,\gamma), & S_1\in [\underline S_1,\overline S_1]\ \textnormal{a.s.},\\
                +\infty, & \textnormal{otherwise}.
          \end{array}\right.
$$                 

Furthermore, the function
$$f(\omega,x)=U\left(1+\gamma_0 (x-S_0)\right)+\delta(x|[\underline S_1(\omega),\overline S_1(\omega)])$$
is Borel on $[0,1] \times \mathbb R$ and lower semicontinuous in $x$ for each $\omega$. Hence, $f$ is a \emph{normal integrand} (see \cite{EkeTem99}, Chapter VIII, Definition 1.1), uniformly bounded from below. The relaxation of $F$ is given by the formula
$$ \widehat F(S_1)=\int_0^1 \widehat f(\omega,S_1(\omega))\,d\omega,$$
where $\widehat f(\omega,\cdot)$ is the largest \emph{convex} lower semicontinuous minorant of $f(\omega,\cdot)$ for each $\omega$: see \cite{EkeTem99}, Chapter IX, Propositions 1.2 and 2.3 or \cite{Roc76}, \cite{HuPap97} (chapter 2, section 9) for more general results of this sort. Using the concavity of $f(\omega,\cdot)$ on $[\underline S_1(\omega),\overline S_1(\omega)]$, we conclude that $\widehat f(\omega,\cdot)$ is linear on this interval:
$$ \widehat f(\omega,S_1)=f(\omega,\overline S_1) \frac{S_1-\underline S_1}{\overline S_1-\underline S_1}+f(\omega,\underline S_1)\frac{\overline S_1-S_1}{\overline S_1-\underline S_1}+\delta(S_1|[\underline S_1,\overline S_1]).$$
Thus, for $S_1\in [\underline S_1,\overline S_1]$ we have
\begin{eqnarray*}
\widehat \Psi(S,\gamma)=\widehat F(S_1) &= &\mathsf E\biggl(U(1+\gamma_0 (\overline S_1-S_0))\frac{S_1-\underline S_1}{\overline S_1-\underline S_1}\\
&+& U(1+\gamma_0 (\underline S_1-S_0))\frac{\overline S_1-S_1}{\overline S_1-\underline S_1}\biggr).
\end{eqnarray*}

Now assume that the function $U$ is strictly increasing and differentiable and there exists an optimal solution $\gamma_0^*$ of (\ref{eq:3.2}). From Theorem \ref{th:3} it follows that $S^*$ is a generalized shadow price iff $(S^*,\gamma^*)$ is a saddle point of $\widehat\Psi$. From the representation 
\begin{eqnarray*}
\widehat \Psi(S,\gamma)=&= &\mathsf E\biggl(\frac{U(1+\gamma_0 (\overline S_1-S_0))-U(1+\gamma_0 (\underline S_1-S_0))}{\overline S_1-\underline S_1} S_1\\
&+& \frac{\overline S_1 U(1+\gamma_0 (\underline S_1-S_0))-\underline S_1 U(1+\gamma_0 (\overline S_1-S_0))}{\overline S_1-\underline S_1}\biggr).
\end{eqnarray*}
it is clear that the inequality $\widehat\Psi(S^*,\gamma^*)\le \widehat\Psi(S,\gamma^*)$, $S_1\in [\underline S_1,\overline S_1]$ is equivalent to the condition
\begin{equation} \label{eq:3.3}
 S^*_1 I_{\{\gamma_0^*\neq 0\}}=\underline S_1 I_{\{\gamma_0^*>0\}}+\overline S_1 I_{\{\gamma_0^*<0\}}.
\end{equation}

Furthermore, the inequality $\widehat\Psi(S^*,\gamma)\le \widehat\Psi(S^*,\gamma^*)$, $\gamma_0\in\mathbb R$ reduces to the condition 
$$ \frac{\partial\widehat\Psi}{\partial\gamma_0}(S^*,\gamma^*)=0$$
due to the concavity of $\widehat\Psi(S^*,\cdot)$. After elementary calculations we get
$$ \frac{\partial\widehat\Psi}{\partial\gamma_0}(S^*,\gamma^*)=\mathsf E\left((\underline S_1-S_0) U'(1+\gamma_0^*(\underline S_1-S_0))\right)=0\ \ \textnormal{for}\ \ \gamma_0^*>0,$$
$$ \frac{\partial\widehat\Psi}{\partial\gamma_0}(S^*,\gamma^*)=\mathsf E\left((\overline S_1-S_0) U'(1+\gamma_0^*(\overline S_1-S_0))\right)=0\ \ \textnormal{for}\ \ \gamma_0^*<0,$$
$$ \frac{\partial\widehat\Psi}{\partial\gamma_0}(S^*,\gamma^*)=U'(1)\mathsf E(S_1^*-S_0)=0\ \ \textnormal{for}\ \ \gamma_0^*=0.$$
Taking into account (\ref{eq:3.3}), we conclude that the last three equalities are equivalent to the following one:
\begin{equation} \label{eq:3.4}
\frac{\partial\widehat\Psi}{\partial\gamma_0}(S^*,\gamma^*)=\mathsf E\left((S_1^*-S_0^*)U'(1+\gamma_0^*(S_1^*-S_0^*))\right)=0.
\end{equation}
Thus, $(S^*,\gamma^*)$ is a saddle point of $\widehat\Psi$ iff the relations (\ref{eq:3.3}), (\ref{eq:3.4}) hold true.

From this observation it follows that any generalized shadow price is a \emph{shadow price}. Indeed, condition (\ref{eq:3.4}) ensures that $\gamma^*$ is an optimal solution in the frictionless model with the price process $S^*$:
$$ \mathsf E U(1+\gamma_0(S_1^*-S_0^*))\le\mathsf E U(1+\gamma_0^*(S_1^*-S_0^*)).$$
Moreover, in view of (\ref{eq:3.3}) we have
$$\mathsf E U(1+\gamma_0^*(S_1^*-S_0^*))=\mathsf E U(1+(\gamma_0^*)^+ (\underline S_1-S_0)-(\gamma_0^*)^- (\overline S_1-S_0))=\widehat\Psi(S^*,\gamma^*)=\lambda.$$

Hence, although 
$$\widehat\Psi(S,\gamma)\neq\Psi(S,\gamma)=\mathsf EU(1+\gamma_0 (S_1-S_0))$$ 
(if, e.g., $U$ is strictly concave), in this example a process $S^*$ is a generalized shadow price iff it is a shadow price. 

\emph{Example 4.}
Let $\Omega=\mathbb N$, $T=1$, $\mathcal F_0$ is generated by the atoms $D_n=\{2n-1,2n\}$, $n\in\mathbb N$ and $\mathcal F_1$ coincides with the power set of $\mathbb N$. The probability measure is defined by $\mathsf P(\{n\})=2^{-n}$. Put
$\underline S_0=1$, $\overline S_0=4$,
$$ \underline S_1=\overline S_1=S_1=\sum_{n=1}^\infty\left (4 I_{\{2n\}}+I_{\{2n-1\}}\right).$$
Since $S_t$ is bounded we can put $s=q=r=\infty$. From the definitions of $X_1(\gamma)$ and $S$ it follows that 
$$ X_1(\gamma)=1+\gamma_0^+ (S_1-\overline S_0)-\gamma_0^- (S_1-\underline S_0)\le 1.$$ 
and $\gamma_0=0$ is an optimal trading strategy for any monotone functional $\Phi$ on $L^\infty(\mathcal F_1)$. 

Denote by $\LIM:L^\infty(\mathcal F_1)\mapsto\mathbb R$ a Banach limit (see e.g. \cite{DunSch58}, Chapter II, Exercise 22) and put
$$ \Phi(X)=\mathsf E X+\LIM(X).$$
Clearly, $\Phi$ is a linear monotone functional on $L^\infty(\mathcal F_1)$. We have 
$$\lambda=\sup\{\Phi(X_1(\gamma)):\gamma_0\in L^\infty(\mathcal F_0)\}=1.$$

We show that there is \emph{no shadow price} in this model. Assume first that $S_0$ is a shadow price which is not equal to the conditional expectation
$$ \mathsf E(S_1|\mathcal F_0)=\sum_{n=1}^\infty\frac{\mathsf E(S_1 I_{D_n})}{\mathsf P(D_n)} I_{D_n}=
\sum_{n=1}^\infty\frac{4\mathsf P(2n)+\mathsf P(2n-1)}{\mathsf P(2n)+\mathsf P(2n-1)}I_{D_n}=2.$$
If $S_0\neq 2$ on $D_n$, then putting $\gamma_0(i)=0$, $i\not\in D_n$; $\gamma_0(i)=\delta$, $i\in D_n$ we get
$$ \Phi(1+(\gamma\circ S)_1)=\mathsf E(1+\gamma_0\Delta S_1)=1+\mathsf E\left(\gamma_0\left(2-S_0\right)\right)=1+\delta\left(2-S_0\right)\mathsf P(D_n).$$
It follows that
$$ \mu_S=\sup\{\Phi(1+(\gamma\circ S)_1):\gamma_0\in L^\infty(\mathcal F_0)\}=+\infty.$$

Now assume that $S_0=\mathsf E(S_1|\mathcal F_0)=2$. For $\gamma_0=1$ we have
$$ \Phi(1+(\gamma\circ S)_1)=\LIM (1+\Delta S_1)=\frac{3}{2}.$$
For computing the value of the Banach limit in the last equality we have used its shift-invariance property: 
\begin{eqnarray*}
2\LIM(\Delta S_1) &=& \LIM\sum_{n=1}^\infty(2 I_{\{2n\}}-I_{\{2n-1\}})+\LIM\sum_{n=1}^\infty (2 I_{\{2n-1\}}-I_{\{2n\}})\\
&=& \LIM (1)=1.
\end{eqnarray*}
Thus, $\mu_S\ge 3/2>\lambda=1$ and $S_0=2$ is not a shadow price. 

The existence of a generalized shadow price $S^*$ is guaranteed by Theorem \ref{th:1}. Let us show that $S^*_0=2$. For $S_0=2$ we have
$$  \Psi(S,\gamma)=\Phi(1+(\gamma\circ S)_1)=\LIM(1+\gamma_0(S_1-S_0)).$$
Consider a neighbourhood $U$ of $S_0$ in the topology $\tau^w_0=\sigma(L^\infty(\mathcal F_0),L^1(\mathcal F_0))$, restricted to $[\underline S_0,\overline S_0]$:
$$ U=\{S_0'\in [\underline S_0,\overline S_0]:|\mathsf E g_i(S_0'-S_0)|<\varepsilon,\ i=1,\dots,m\},\ \ \ 
g_i\in L^1(\mathcal F_0),\ \varepsilon>0.$$
The set 
$$U_n=\left\{S_0'\in [\underline S_0,\overline S_0]: S_0'=S_0\ \textnormal{on}\ \bigcup_{j=1}^n D_j\right\}$$ %\{S_0^1\}\times\dots\times\{S_0^{n-1}\}\times\prod_{j\ge n} [\underline S_0^j,\overline S_0^j]$$
is contained in $U$ for sufficiently large $n$. Indeed, for $S'_0\in U_n$ we have
$$ |\mathsf E g_i(S_0'-S_0)|\le\sum_{j=n+1}^\infty |g_i^j| |\overline S_0^j-\underline S_0^j|\mathsf P(D_j),$$
where $g_i^j=g_i(\omega)$, $\omega\in D_j$ and $\overline S_0^j$, $\underline S_0^j$ are defined similarly.
The right-hand side of the last inequality can be made arbitrary small by an appropriate choice of $n$. 

Take $S'_0\in U_n\subset U$ such that 
$$1+\gamma_0\Delta S'_1=1+\gamma_0^+ (S_1-\overline S_0)-\gamma_0^- (S_1-\underline S_0)=X_1(\gamma)\ \ \ \textnormal{on}\ \bigcup_{j\ge n+1} D_j.$$
Clearly, 
$$\Psi(S',\gamma)=1+\LIM (\gamma_0\Delta S'_1)=\LIM(X_1(\gamma)).$$ 
It follows that $\widehat\Psi(S,\gamma)\le\LIM(X_1(\gamma))\le 1$ and $S_0^*=2$ determines a generalized shadow price.

\emph{Example 5.} Let us reproduce the counterexample of \cite{BenCamKalMuh11}. Put $\Omega=\mathbb N$, $T=2$, $\mathcal F_0=\{\emptyset,\Omega\}$. Let $\mathcal F_1$ be generated generated by the atoms 
$$D_k=\{2k+1,2k+2\},\ \ \ k\in\mathbb N_0:=\{0\}\cup\mathbb N,$$ 
and let $\mathcal F_2$ be the power set of $\mathbb N$. 

Assume that the stock bid prices are falling deterministically: $\underline S_0=3$, $\underline S_1=2$ $\underline S_2=1$ and the ask prices are defined as follows: $\overline S_0=3$, 
$$ \overline S_1=2+k\ \textnormal{on}\ D_k,\ \ k\in\mathbb N_0, $$
$$ \overline S_2(\omega)=1\ \ \textnormal{for}\ \omega=2k+1,\ \ k\in\mathbb N_0,$$ 
$$ \overline S_2(\omega)=3+k\ \ \textnormal{for}\ \omega=2k+2,\ \ k\in\mathbb N_0.$$ 

The probability measure is defined as follows:
$$ \mathsf P(D_0)=1-2^{-n},\ \ \ \mathsf P(D_k)=2^{-n-k};$$
%$$ \mathsf P(\{1\})=(1-2^{-n})\mathsf P(D_0),\ \ \mathsf P(\{2\})=2^{-n}\mathsf P(D_0),$$
$$ \mathsf P(\{2k+1\})=(1-2^{-n-k})\mathsf P(D_k),\ \  \mathsf P(\{2k+2\})=2^{-n-k}\mathsf P(D_k),\ \ k\in\mathbb N_0,$$
where $n\in\mathbb N$ is fixed sufficiently large to make $\mathsf E(\overline S_2-\underline S_1|\mathcal F_1)<0$.

The problem cosists in maximization of the logarithmic utility $\Phi(X_2(\gamma))=\mathsf E\ln(X_2(\gamma))$ (we put $\ln x=-\infty$ for $x\le 0$). Expectation is defined by the formula  
$$ \mathsf E f=\lim_{M\to+\infty}\mathsf E(f\wedge M).$$
for any measurable function $f$ with values in the extended real line $\mathbb R\cup\{\pm\infty\}$.
Particularly, $\mathsf E f=-\infty$ if $\mathsf E f^-=+\infty$.

The picture and clear economical argumentation, given in \cite{BenCamKalMuh11}, show that it is optimal not to trade at step $0$ ($\gamma^*_0=0$) and to go short at step $1$ ($\gamma^*_1<0$). To be a bit more formal consider
\begin{eqnarray} \label{eq:3.5}
X_2(\gamma)&=& 1+\sum_{t=0}^2 \left(\underline S_t (\Delta\gamma_t)^- - \overline S_t (\Delta\gamma_t)^+\right)\nonumber\\
           &=& 1-S_0\gamma_0+\underline S_1(\gamma_1-\gamma_0)^{-}-\overline S_1(\gamma_1-\gamma_0)^+ + \underline S_2\gamma_1^+ - \overline S_2 \gamma_1^{-},
\end{eqnarray}           
where $S_0=\underline S_0=\overline S_0$. It is easy to see that $\gamma_0<0$ leads to a negative value of $X_2(\gamma)$ for some $\omega=2k$. Assuming that $\gamma_0\ge 0$, it is not optimal to posses a positive amount of stock at step 1:
$$ X_2(\gamma)=1+(\underline S_1-S_0)\gamma_0\ \ \textnormal{for}\ \gamma_1=0;$$
$$ X_2(\gamma)=1+(\underline S_1-S_0)\gamma_0+(\underline S_2-\underline S_1)\gamma_1\le 1+(\underline S_1-S_0)\gamma_0\ \ \textnormal{for}\ \gamma_1\in (0,\gamma_0);$$
\begin{eqnarray*}
X_2(\gamma) &=& 1+(\overline S_1-S_0)\gamma_0+(\underline S_2-\overline S_1)\gamma_1\le 
1+(\underline S_1-S_0)\gamma_0\ \ \textnormal{for}\ \gamma_1\ge \gamma_0,
\end{eqnarray*}
since $ (\underline S_2-\overline S_1)\gamma_1\le (\underline S_1-\overline S_1)\gamma_1\le (\underline S_1-\overline S_1)\gamma_0$ for $\gamma_1\ge \gamma_0$.

Under the assumptions $\gamma_0\ge 0$, $\gamma_1\le 0$ the expression (\ref{eq:3.5}) reduces to
$$ X_2(\gamma)=1+(\underline S_1-S_0)\gamma_0+(\overline S_2-\underline S_1)\gamma_1\le 1+(\overline S_2-\underline S_1)\gamma_1.$$
It follows that the maximization of $\mathsf E\ln(X_2(\gamma))$ can be carried over the set $\{\gamma_0=0, \gamma_1\le 0\}$:
\begin{eqnarray} \label{eq:3.6}
\lambda &=& \sup\{\mathsf E\ln(X_2(\gamma)):\gamma_t\in L^0(\mathcal F_t),\ t=0,1\}\nonumber\\
       &=&\sup\{\mathsf E\ln(1+(\overline S_2-\underline S_1)\gamma_1):\gamma_1\in L^0(\mathcal F_1,-\mathbb R_+)\}.
\end{eqnarray}

Moreover, since $\mathsf E(\overline S_2-\underline S_1|\mathcal F_1)<0$, it is not optimal to do nothing.  Denote by $\gamma_1^k$ the value of $\gamma_1$ on $D_k$. We have
\begin{eqnarray} \label{eq:3.7}
\mathsf E\ln(1+(\overline S_2-\underline S_1)\gamma_1) &=& \sum_{k=0}^\infty\biggl((1-2^{-n-k})\ln(1-\gamma_1^k)\nonumber\\
&+ &     2^{-n-k}\ln\left(1+(1+k)\gamma_1^k\right)\biggr)\mathsf P(D_k).
\end{eqnarray}
The optimal portfolio $\gamma_1^{*,k}<0$ can be obtained by maximizing each term in this sum. Taking into account that 
$\gamma_1^{*,k}\in (-(1+k)^{-1},0)$, we get an estimate 
\begin{eqnarray*}
(1-2^{-n-k})\ln(1-\gamma_1^k) + 2^{-n-k}\ln\left(1+(1+k)\gamma_1^k\right)\\
\le  (1-2^{-n-k}) (-\gamma_1^k) \le \frac{1-2^{-n-k}}{k+1}
\end{eqnarray*}   
which shows that the optimal utility value $\lambda$ is finite.

Since the optimal strategy $\gamma_1^*<0$, $\gamma_2^*=-\gamma_1^*$ is active, shadow prices $S_1^*$, $S_2^*$ should coincide with $\underline S_1$, $\overline S_2$. Otherwise, the same strategy would give strictly higher utility value in the frictionless market with stock price $S^*$. But in this frictionless market the optimal utility value $\mu_{S^*}$ is infinite since
$$1+\gamma_0\Delta S_1^*=1-\gamma_0\to +\infty,\ \ \gamma_0\to-\infty.$$   
Thus, there is \emph{no shadow price} in this model.

However, as we will see shortly, the process $S^*=(S_0,\underline S_1,\overline S_2)$ is a \emph{generalized} shadow price. The point is that in the relaxed problem short selling at step $0$ is automatically prohibited. Put $s=0$, $q=1$ in the notation of Section \ref{sec:2}. By the same reasons as in Example 2, the topology $\tau^p_t$ of pointwise convergence coincides with the weak topology $\tau_t^w=\sigma(L^1(\mathcal F_t),L^\infty(\mathcal F_t))$ on the set $[\underline S_t,\overline S_t]$. For any $\prod_{t=0}^2\tau^p_t$-neighbourhood $U$ of $S^*$ there exist sufficiently large $k$ and $S'\in U$ such that $S_1'=\overline S_1$, $S_2'=\overline S_2$ on $D_k$. We have
\begin{eqnarray*}
 1+(\gamma\circ S')_2 &=& 1+(\overline S_1-S_0)\gamma_0+(\overline S_2-\overline S_1)\gamma_1\\
 &=& (k-1)\gamma_0+(I_{\{2k+2\}}-(k+1)I_{\{2k+1\}})\gamma_1^k\ \ \ \textnormal{on}\ D_k.
\end{eqnarray*} 
If $\gamma_0<0$, then $1+(\gamma\circ S')_2(2k+2)<0$ for large $k$. Thus, 
$$\Psi(S',\gamma)=\mathsf E\ln(1+\gamma\circ S')_2=-\infty$$ 
and $\widehat\Psi(S^*,\gamma)=-\infty$ for any $\gamma_0<0$. 

Furthermore,
$$ 1+(\gamma\circ S^*)_2=1+(\underline S_1-S_0)\gamma_0+(\overline S_2-\underline S_1)\gamma_1\le 1+(\overline S_2-\underline S_1)\gamma_1\ \ \textnormal{for}\ \gamma_0\ge 0.$$
It follows that $\widehat\Psi(S^*,(\gamma_0,\gamma_1))\le\widehat\Psi(S^*,(0,\gamma_1))$ and one can assume $\gamma_0=0$ in the relaxed utility maximization problem (\ref{eq:2.6}):
$$\widehat\mu_{S^*} =\sup\{\widehat\Psi(S^*,\gamma):\gamma_0=0, \gamma_1\in L^0(\mathcal F_1)\}.$$
To prove that $\mu_{S^*}=\lambda$, we go back to the ''unrelaxed'' frictionless problem:
\begin{eqnarray*}
\widehat\mu_{S^*} &\le & \sup\{\Psi(S^*,\gamma):\gamma_0=0, \gamma_1\in L^0(\mathcal F_1\}\\
 &=& \sup\{\mathsf E\ln\left(1+(\overline S_2-\underline S_1)\gamma_1\right):\gamma_1\in L^0(\mathcal F_1)\}.
\end{eqnarray*}
Looking again at (\ref{eq:3.7}), we conclude that optimal values $\gamma_1^{*,k}$ are negative. Comparing the last expression with (\ref{eq:3.6}), we obtain the inequality $\widehat\mu_{S^*}\le\lambda$. The reverse inequality is evident.

\section{Shadow prices via Lagrange duality} \label{sec:4}
\setcounter{equation}{0}
As is known, in mathematical economics
%From the mathematical and economical literature it is known that 
shadow resource prices are associated with the optimal solution of the dual problem: see e.g. \cite{Koo76}, \cite{Fia83} (Chapter 5). To avoid conflicts with the terminology of the present paper we, following \cite{Roc70}, use the term ''equilibrium prices'' instead. These prices are introduced along the following lines. 
Let $x=(x_1,\dots,x_n)$ represent activities of a firm and let $f(x)$ be the cost of the corresponding operation. The activities are subject to the resource constraints $g_i(x)\le 0$, $i=1,\dots,m$. Put
$$ \varphi(u)=\inf\{f(x):g_i(x)\le u_i,\ i=1,\dots,m\}.$$
The components of a vector $\lambda^*=(\lambda_1^*,\dots,\lambda_m^*)$ are called \emph{equilibrium} resource prices if the firm cannot reduce the optimal cost of the operation by buying or selling resources at these prices:
$$ \varphi(u)+\sum_{i=1}^m \lambda_i^* u_i\ge\varphi(0),\ \ \ u\in\mathbb R^m.$$
For a convex problem vectors $\lambda^*$ of equilibrium prices are exactly the optimal solutions of the dual problem (see \cite{Roc70}, Theorem 28.2 and Corollary 28.4.1 for the precise statement). 

In the problem under consideration the trader has two resources at his disposal: bonds and stocks. It is natural to expect that the equilibrium prices of these resources are related to the shadow price process introduced above. 

Assume that $\Omega$ is finite, $\mathcal F_T$ coincide with the power set of $\Omega$ and $\mathsf P(\omega)>0$, $\omega\in\Omega$. First of all we rewrite the self-financing condition, separating the ''resource constraints'':
\begin{equation} \label{eq:4.1}
(\Delta\beta_t-L_t\underline S_t+M_t\overline S_t)(\omega)\le 0,\ \ \ t\in 0,\dots,T,\ \omega\in\Omega;
\end{equation}
\begin{equation} \label{eq:4.2}
(\Delta\gamma_t+L_t-M_t)(\omega)\le 0,\ \ \ t\in 0,\dots,T,\ \omega\in\Omega;
\end{equation}
\begin{equation} \label{eq:4.3}
-L_t(\omega)\le 0,\ \ -M_t(\omega)\le 0,\ \ \ t\in 0,\dots,T,\ \omega\in\Omega.
\end{equation}
Here, as above, $\beta_{-1}=1$, $\gamma_{-1}=0$. By $L_t$ (respectively, $M_t$) we denote the number of stocks sold (respectively, purchased) at time $t$ at price $\underline S_t$ (respectively, $\overline S_t$). Clearly, passing to the inequality constraints (corresponding to the possibility of consumption) and allowing the simultaneous transfers from bonds to stocks and back: $L_t M_t\neq 0$ do not increase trader's monotone utility. We should also take into account 
the ''boundary condition'':
\begin{equation} \label{eq:4.4}
\gamma_T(\omega)=0, \ \ \omega\in\Omega 
\end{equation}
and the ''information constraints'':
\begin{equation} \label{eq:4.5}
(\beta_t,\gamma_t,L_t,M_t)\in L^0(\mathcal F_t,\mathbb R^4),\ \ t\in 0,\dots,T.
\end{equation}

Consider a concave utility function $U$ as in Example 1: $U$ is finite on $(0,\infty)$ and $U(x)=-\infty$, $x\le 0$ 
and denote by $C$ the set of processes $(\beta,\gamma,L,M)$, satisfying (\ref{eq:4.5}) and such that $\beta_T>0$.
The problem is to minimize
\begin{equation} \label{eq:4.6}
 -\mathsf EU(\beta_T)
\end{equation} 
over the set $C$ under the constraints (\ref{eq:4.1}) -- (\ref{eq:4.4}). Formally, this is an \emph{ordinary convex optimization program} (\cite{Roc70}, Section 28).

Consider the Lagrange function
\begin{eqnarray} \label{eq:4.7}
\mathcal L= &-& \mathsf EU(\beta_T)+\sum_{t=0}^T\mathsf E(Z_t^1(\Delta\beta_t-L_t\underline S_t+M_t\overline S_t))+\sum_{t=0}^T\mathsf E(Z_t^2(\Delta\gamma_t+L_t-M_t))\nonumber\\
&-& \sum_{t=0}^T\mathsf E(Z_t^3 L_t)-\sum_{t=0}^T\mathsf E(Z_t^4 M_t)+\mathsf E(\nu_T\gamma_T)\ \ \
\textnormal{for}\  (\beta,\gamma,L,M)\in C.
\end{eqnarray}
The Lagrange multipliers are represented by a process $Z_t=(Z_t^1,Z_t^2,Z_t^3,Z_t^4)$ with non-negative components: $Z_t\in L^0(\mathcal F_t,\mathbb R^4_+)$, $t=0,\dots,T$ and $\nu_T\in L^0(\mathcal F_T)$. Note that the process $Z$ may be assumed adapted since for adapted processes $(\beta,\gamma,L,M)\in C$ the number of constraints in (\ref{eq:4.1}) -- (\ref{eq:4.3}) for fixed $t$ coincides with the number of atoms of $\mathcal F_t$ and $Z_t$ can be taken constant on these atoms. 

To complete the definition of $\mathcal L$, we put in accordance to the general scheme of \cite{Roc70} (Section 28)
$$ \mathcal L=+\infty,\ \ \textnormal{if}\ (\beta,\gamma,L,M)\not\in C;$$
$$ \mathcal L=-\infty,\ \ \textnormal{if}\ (\beta,\gamma,L,M)\in C,\ \ Z_t\not\in L^0(\mathcal F_t,\mathbb R^4_+)\ \textnormal{for some}\ t.$$

Collecting terms, containing the same elements $\beta_t$, $\gamma_t$, $L_t$, $M_t$, we rewrite (\ref{eq:4.7}) in the following way:
$$\mathcal L=\mathcal L_1+\mathcal L_2+\mathcal L_3,$$
$$\mathcal L_1=\mathsf E\left(-U(\beta_T)+Z_T^1\beta_T\right)-\sum_{t=0}^{T-1}\mathsf E\beta_t\Delta Z_{t+1}^1-\mathsf E Z_0^1,$$
$$\mathcal L_2=\mathsf E\gamma_T(Z_T^2+\nu_T)-\sum_{t=0}^{T-1}\mathsf E\gamma_t\Delta Z_{t+1}^2,$$
$$ \mathcal L_3=\sum_{t=0}^T\mathsf E L_t(Z_t^2-Z_t^1\underline S_t-Z_t^3)+\sum_{t=0}^T\mathsf E M_t(-Z_t^2+Z_t^1\overline S_t-Z_t^4).$$

The objective function of the dual problem is given by
$$g(Z,\nu_T)=\inf\{\mathcal L:(\beta,\gamma,L,M)\in C\}. $$
Put $V(x)=\inf\limits_{y}(-U(y)+xy)$. After simple calculations we get
$$ g(Z,\nu_T)=\mathsf E V(Z_T^1)-\mathsf E Z_0^1 $$
if $Z_t\in L^0(\mathcal F_t,\mathbb R^4_+)$, $t=0,\dots,T$ and the following conditions hold true
$$ \mathsf E(\Delta Z^1_{t+1}|\mathcal F_t)=0,\ \ \mathsf E(\Delta Z^2_{t+1}|\mathcal F_t)=0,\ \ t=0,\dots,T-1;$$
$$ Z_t^2-Z_t^1\underline S_t=Z_t^3,\ \ \ -Z_t^2+Z_t^1\overline S_t=Z_t^4, \ \ t=0,\dots,T;\ \ \nu_T=-Z_T^2.$$
Otherwise, $g(Z,\nu_T)=-\infty$. 

It readily follows that the optimal value of the dual problem can be represented as
\begin{equation} \label{eq:4.8}
 \sup\{\mathsf E (V(Z_T^1)-Z_0^1): Z\in D\},
\end{equation}
$$ D=\{Z^1\in\mathcal M_+: Z_t^1\underline S_t\le Z_t^2\le Z_t^1\overline S_t\ \textnormal{for some}\ Z^2\in\mathcal M_+\},$$
where $\mathcal M_+$ is the set of non-negative $\mathsf P$-martingales. The representations of this sort are well known: see \cite{CviKar96} for continuous time case and \cite{Pen11}, \cite{PenPer11} for generalizations in discrete time.

The objective function (\ref{eq:4.6}) of the primal problem is finite on $C$ and the point $(\overline\beta,\overline\gamma)$, where $\overline\beta_t=1$, $\overline\gamma_t=0$, $t=0,\dots T$, belongs to the relative interior of $C$ and satisfies the constraints (\ref{eq:4.1}) -- (\ref{eq:4.4}), which are affine. If the optimal value $-\lambda$ of the primal problem is finite then there is no duality gap and the dual problem is solvable (\cite{Roc70}, Theorem 28.2 and Corollary 28.4.1). That is, 
$$ -\lambda=\sup\{\mathsf E (V(Z_T^1)-Z_0^1): Z\in D\}=\mathsf E (V(\widehat Z_T^1)-\widehat Z_0^1)$$
fo some $\widehat Z^1\in D$.

Let us introduce an adapted process $S_t^*\in [\underline S_t,\overline S_t]$ such that 
\begin{equation} \label{eq:4.9}
S_t^* \widehat Z_t^1=\widehat Z_t^2.
\end{equation}
On the atoms of $\mathcal F_t$ with $\widehat Z_t^1=0,\widehat Z_t^2=0$ the values $S_t^*\in [\underline S_t,\overline S_t]$ are chosen arbitrary. Put
\begin{equation} \label{eq:4.10}
-\mu_{S^*}=\sup\{\mathsf E (V(Z_T^1)-Z_0^1): Z\in D(S^*)\},
\end{equation}
$$ D(S^*)=\{Z^1\in\mathcal M_+: Z^2=Z^1 S^*\in\mathcal M_+\}.$$
The maximization in (\ref{eq:4.10}) is carried over smaller set as compared to (\ref{eq:4.8}), and the objective functions are the same. Hence, $-\lambda\ge -\mu_{S^*}$. On the other hand, the optimal solution $\widehat Z$ of (\ref{eq:4.8}) is feasible for (\ref{eq:4.10}): $\widehat Z^1\in D(S^*)$ since $\widehat Z^2=S_t^* \widehat Z_t^1\in\mathcal M_+$. It follows that $\lambda=\mu_{S^*}$. But (\ref{eq:4.10}) is the dual to the frictionless optimization problem with the price process $S^*$. This means that $S^*$ is a shadow price in the sense of the definition of Section \ref{sec:2}.

In fact, we have obtained the same result as in Example 1 (and in \cite{KalMuh-Kar11}). But formula (\ref{eq:4.9}) reveals one more interpretation of a shadow price process: it is the equilibrium bond/stock exchange rate, that is, the relation of stock and bond equilibrium prices.

\end{document}